\documentclass[superscriptaddress,showpacs,aps,twocolumn]{revtex4}
\usepackage{amssymb}
\usepackage{amsmath}
\usepackage{graphicx}
\usepackage{color}
\usepackage{amsfonts}
\usepackage{multirow}
\usepackage{tabularx}

\begin{document}

\title{The interaction of atoms with LiF(001) revisited}
\author{J.E. Miraglia}
\affiliation{Instituto de Astronom\'{\i}a y F\'{\i}sica del Espacio (IAFE, CONICET-UBA),
casilla de correo 67, sucursal 28, C1428EGA, Buenos Aires, Argentina.}
\affiliation{Depto. de F\'{\i}sica, Fac. de Ciencias Exactas y Naturales, Universidad de
Buenos Aires, Argentina.}
\author{M. S. Gravielle}
\affiliation{Instituto de Astronom\'{\i}a y F\'{\i}sica del Espacio (IAFE, CONICET-UBA),
casilla de correo 67, sucursal 28, C1428EGA, Buenos Aires, Argentina.}
\date{\today }

\begin{abstract}
Pairwise additive potentials for multielectronic atoms interacting with a
LiF(001) surface are revisited by including an improved description of the
electron density associated with the different lattice sites, as well as
non-local electron density contributions. Within this model, the electron
distribution around each ionic site of the crystal is described by means of
an \textit{onion} approach that accounts for the influence of the Madelung
potential. From such densities, binary interatomic potentials are then
derived by using well-known non-local functionals for the kinetic, exchange
and correlation terms. Rumpling and long-range contributions due to
projectile polarization and van der Waals forces are also included in an
analogous fashion. We apply this pairwise additive approximation to evaluate
the interaction potential between closed-shell - He, Ne, Ar, Kr, and Xe -
and open-shell \ - N, S, and Cl - atoms and the LiF surface, analyzing the
relative importance of the different contributions. The performance of the
proposed potentials is assessed by contrasting angular positions of rainbow
and supernumerary rainbow maxima produced by fast grazing incidence with
available experimental data. The good agreement found for normal energies in
the eV- range represents a meaningful evidence of the quality of the present
description.
\end{abstract}

\date{\today}
 \maketitle

\section{INTRODUCTION}

In the field of particle-surface interactions, one of the most remarkable
experimental advances of the last decade corresponds to the observation of
grazing incidence fast atom diffraction (GIFAD or FAD) \cite%
{Schuller2007,Rousseau2007}, which has emerged as a powerful surface
analysis technique \cite{Khemliche2009,Winter2011,Atkinson2014,Seifert2016}.
It allows one to inspect the electronic and morphological characteristics of
crystal surfaces with an exceptional sensitivity, becoming a useful tool for
investigating a wide variety of materials \cite%
{Schuller2009b,Seifert2010,Seifert2013,Zugarramurdi2015}.

The accuracy of the surface information provided by the FAD method crucially
relies on the theoretical model used to describe the surface potential. In
previous articles \cite{Gravielle2008,Gravielle2009,Gravielle2011} we
investigated the FAD process for different atoms impinging on LiF(001) by
using a pairwise additive approach to represent the surface interaction.
Pairwise additive potentials are built as a sum of binary interatomic
potentials that describe the interaction of the atomic projectile with
individual ionic centers of the crystal. For insulator materials, like LiF,
where the electron density is highly localized around the atomic nuclei,
this simple potential model has shown to represent a reliable alternative to
more complex self-consistent \textit{ab-initio} calculations \cite%
{Schuller2008b,Schuller2009d,Gravielle2009,Specht2011,Gravielle2011}.
However, in most works the binary potentials\ are derived by using the\
Local Density Approximation (LDA) to evaluate\ the kinetic and exchange
terms \cite{Abrahamson}. But the LDA does not include contributions due to
non-local electron density terms, which might play an important role,
especially for open-shell projectiles, with partially occupied outer shells.

In Ref. \cite{Gravielle2013} an attempt to include the proper asymptotic
limit of the binary interatomic potentials was done for the case of
multielectronic atoms grazingly scattered off a LiF(001) surface. In this
article we revisit such pairwise additive model by incorporating non-local
contributions of the electron density, together with the improvement of the
description of the electron density associated with each ionic center of the
insulator.

The interaction between rare gases (closed-shell atoms) - He, Ne, Ar, Kr,
and Xe - with fully occupied valence shells, as well as open-shell \ atoms -
N, S, and Cl - with vacancies in the outer level, \ and a LiF(001) surface
is studied. In all the cases, the kinetic, exchange and correlation terms of
the binary potentials are evaluated from well established non-local
functionals, which depend on first- and second- order derivatives of the
electron density. The electron density corresponding to each ionic center\
of the LiF crystal is obtained from an \textit{onion} model that takes into
account the influence of the whole crystal lattice, i.e., the Madelung
potential \cite{Miraglia2011}. Long-range contributions associated with
polarization and van der Waals (vdW) forces, produced by the rearrangement
of the atom and surface densities as a result of the mutual interaction, are
determined within a similar pairwise additive scheme. Furthermore, a surface
rumpling with a displacement distance extracted from \textit{ab-initio}
calculations \cite{Schuller2009} is considered.

With the aim of testing the proposed potential model, we use it to evaluate
angular distributions of fast atoms\ grazingly scattered from the LiF
surface along low-indexed crystallographic channels. The elastic collision
process is described within the surface initial-value representation (SIVR)
approximation \cite{Gravielle2014,Gravielle2015,Gravielle2016}, which is a
semiquantum method that offers a very good representation of the diffraction
spectra, without requiring the use of convolutions to smooth the theoretical
curves \cite{Rubiano2013}. The validity of the surface potential model is
examined by comparing the angular positions of rainbow and supernumerary
rainbow maxima with available experimental data. The rainbow peak
corresponds to the outermost maximum of the projectile distribution, which
has a classical origin, while supernumerary rainbows are associated with
FAD, being produced by quantum interference inside the channel. These
deflection angles are extremely sensitive to the corrugation of the surface
potential across the incidence direction, resulting a useful magnitude to
probe surface interactions.

\

The article is organized as follows. The constituent parts of the binary
interatomic potentials are summarized in Sec. II. In this section we also
show the short-range binary potentials for the different atomic projectiles
- He, N, Ne, S, Ar, Cl, Kr and Xe - interacting with LiF(001), examining
their asymptotic limits. In Sec. III we derive the corresponding total
atom-surface potentials, including projectile polarization and vdW
contributions. In Sec. IV, angular positions of rainbow and supernumerary
rainbow maxima produced by projectile incidence along the $\langle
110\rangle $ and \ $\langle 100\rangle $ channels of the LiF crystal are
compared with experimental data in order to illustrate the soundness of the
potential model. In such a comparison, normal energies, associated with the
projectile motion perpendicular to the axial channel, in the range from $0.2$
to $80$ eV are considered. In Sec. V we outline our conclusions. Atomic
units ($e^{2}=\hbar =m_{e}=1$) are used unless otherwise stated.\medskip\

\section{BINARY \ INTERACTION\ MODEL}

Within a pairwise additive scheme, the interaction between \ an impinging
atom\ and an ionic crystal surface, like LiF(001), is described as a sum of
binary interatomic potentials which depend on the electron densities
corresponding to the atom and individual ionic centers of the crystal. In
this section we will summarize all steps required to build our binary
interatomic potentials, analyzing separately the asymptotic limits of each
contribution.

\subsection{Ionic centers of the crystal: \textit{Onions}}

\begin{figure}[tbp]
\includegraphics[width=0.5\textwidth]{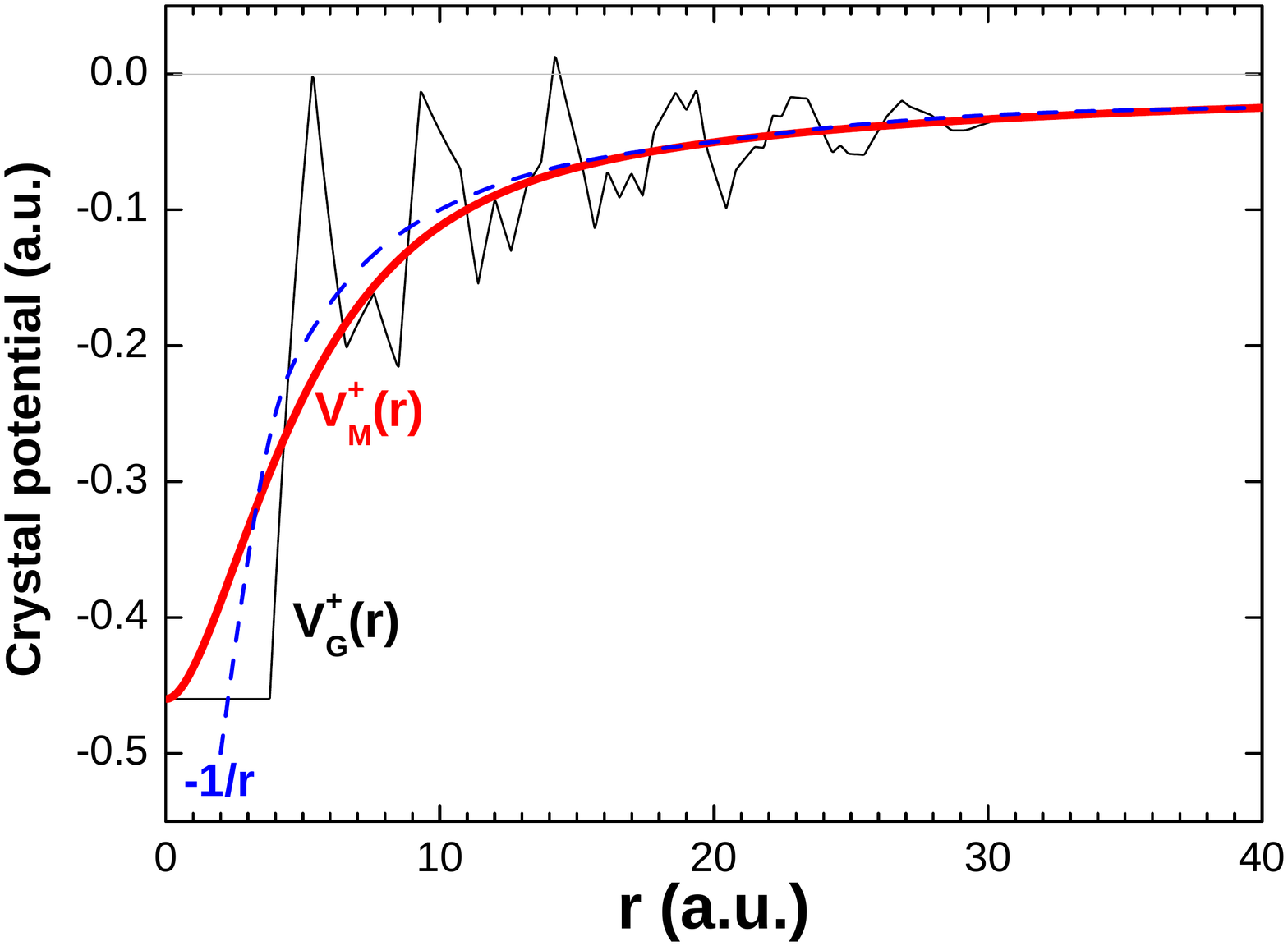}
\caption{(Color online) Crystal potential, as a function of the radial
distance $r$ to an F$^{-}$ site of the crystal lattice. Red thick solid
line, Madelung potential, $V_{M}^{+}$, as given by Eq. (\protect\ref{50});\
black thin solid line, radial grid \ potential, $V_{G}^{+}$, as given by Eq.
(6) of Ref. \protect\cite{Miraglia2011}; blue dashed line, asymptotic limit $%
-1/r$ of the crystal potential.}
\label{Madelung}
\end{figure}

With the purpose of determining the electron density corresponding to each
ionic center of the LiF crystal, let us consider a perfect cubic piece of
crystal centered on an active fluor ion. For this F$^{-}$ anion, immersed in
the LiF crystal, the corresponding multielectronic wave function $\Psi _{%
\mathrm{F}}$ can be obtained by solving the Schr\"{o}dinger equation
associated with the Hamiltonian%
\begin{equation}
H_{\mathrm{F}}=\sum\limits_{l=1}^{10}\left( -\frac{1}{2}\nabla _{\mathbf{r}%
_{_{l}}}^{2}-\frac{Z_{\mathrm{F}}}{r_{l}}+V_{G}^{+}(\mathbf{r}%
_{_{l}})\right) +\frac{1}{2}\sum\limits_{\substack{ k,l=1  \\ k\neq l}}^{10}%
\frac{1}{r_{kl}},  \label{10a}
\end{equation}%
where $\mathbf{r}_{_{l}}$ is the position vector of the $l$- electron ($%
l=1,...,10$) with respect to the F$^{-}$ nuclear charge $Z_{\mathrm{F}}=9$, $%
r_{kl}=\left\vert \mathbf{r}_{k}-\mathbf{r}_{_{l}}\right\vert $ is the
interelectronic distance, and $V_{G}^{+}$ is the potential created by the
whole crystal grid, excluding the active F$^{-}$ ion. In \ Eqs. (6) and (11)
of Ref. \cite{Miraglia2011} $V_{G}^{+}$ was approximated by a radial \textit{%
onion} potential produced by a series of charged shells. But these discrete
charged layers introduce structures in the potential, which are difficult to
handle. Therefore, in this work we fit the previous grid potential \cite%
{Miraglia2011} by means of a smooth potential, here named Madelung
potential, defined as%
\begin{eqnarray}
V_{G}^{+}(\mathbf{r}) &\simeq &V_{M}^{+}(r)=-\frac{1}{r}+\frac{e^{-r/\lambda
}}{r}\left[ 1+(1-\lambda V_{M0})\frac{r}{\lambda }\right.  \notag \\
&+&\left. (\frac{1}{2}-\lambda V_{M0})\left( \frac{r}{\lambda }\right) ^{2}%
\right] ,  \label{50}
\end{eqnarray}%
where $V_{M0}=0.4600$ a.u. represents the proper Madelung potential at the
origin \cite{tatewaki95} and the parameter $\lambda $ is chosen to verify
that the spacial integral of $V_{M}^{+}(r)$ in the range $(0,+\infty )$
coincides with the one of the grid potential of Ref. \cite{Miraglia2011},
finding $\lambda \simeq 0.3\ a/2$ , with $a=7.60$ \ a.u. being the lattice
constant. In Fig. \ref{Madelung} the potential $V_{M}^{+}(r)$ is plotted
along with the grid potential of Ref. \cite{Miraglia2011}. Notice $\ $that $%
V_{M}^{+}(r)$ yields the correct asymptotic limit at long distances, i.e., $%
V_{M}^{+}(r)\rightarrow -1/r$ as $r\rightarrow +\infty $, reproducing the
Coulomb potential originated by an unitary charge which renders the Coulomb
hole that the electron leaves behind when it is removed.

In a similar way,\ the two-electron wave function $\Psi _{\mathrm{Li}}$
corresponding to an active Li$^{+}$ cation, immersed in the LiF crystal, can
be derived within an \textit{onion} treatment from the approximate
Hamiltonian%
\begin{equation}
H_{\mathrm{Li}}\cong \sum\limits_{l=1}^{2}\left( -\frac{1}{2}\nabla _{%
\mathbf{r}_{_{l}}}^{2}-\frac{Z_{\mathrm{Li}}}{r_{l}}+V_{M}^{-}(r_{l})\right)
+\frac{1}{2}\sum\limits_{\substack{ k,l=1  \\ k\neq l}}^{2}\frac{1}{r_{kl}},
\label{10b}
\end{equation}%
where $Z_{\mathrm{Li}}=3$ is the Li$^{+}$ nuclear charge and $%
V_{M}^{-}(r)=-V_{M}^{+}(r)$.

For convenience, we call \textit{onions} to these dressed anion and cation,
denoting them as $F_{@}^{-}\ $\ and $Li_{@}^{+}$, respectively. The
unperturbed electron density associated with each individual \textit{onion}
- $F_{@}^{-}\ $\ or $Li_{@}^{+}$ - is obtained from the square modulus of
the corresponding wave function, $\Psi _{\mathrm{F}}$ or $\Psi _{\mathrm{Li}%
} $, respectively. To evaluate the electronic wave functions $\Psi _{\mathrm{%
F}}$ and $\Psi _{\mathrm{Li}}$ we carried out \textit{\ full} Hartree-Fock
(HF) calculations from the Hamiltonians of Eqs. (\ref{10a}) (with $V_{M}^{+}$
$\ $instead of $V_{G}^{+}\ $) and (\ref{10b}), respectively, using the code
NRHF by Johnson \cite{wjohnson}. The original code was adapted to
incorporate the central potential $V_{M}^{\pm }(r)$ and a grid of about 10$%
^{3}\ $points was used in the numerical calculation.

Binding energies and mean radii derived from the electronic wave functions $%
\Psi _{\mathrm{F}}$ and $\Psi _{\mathrm{Li}}$ are tabulated in Table I. \
From this table, the binding energy of the $F_{@}^{-}(2p)$ is about $-15$ eV
($-0.553$ a.u.), agreeing fairly well with the experimental finding of $(-13$
$\pm 0.3)$ eV for the center of the surface valence band, measured with
respect to the vacuum level \cite{roncin99} .\ Also the ionization energies
of $F_{@}^{-}(2s)$ and $Li_{@}^{+}(1s)$ are near to the experimental values
\cite{samarin04}, while the ionization energy of the inner state $%
F_{@}^{-}(1s)$ is very close to the value obtained as the energy of the
isolated F$^{-}(1s)$\ \cite{Clementi-Roetti}\ minus $V_{M0}$.

\subsection{Binary interatomic potentials}

Making use of the electron densities derived within the \textit{onion} model
(Sec. II A), in this subsection we calculate the binary interatomic
potential between an \textit{onion} $O$\ of the crystal surface, with $O=$ $%
F_{@}^{-}\ $or $O=$ $Li_{@}^{+}$,$\ $and\ the impinging atom $A$, as a
function of the internuclear separation $R$. This binary potential is here
split into two terms: one named \ \textit{short-range} that describes the
static interaction between the atomic projectile and the ionic center $O$ by
considering their respective electron distributions as frozen, and the
other, called \textit{long-range}, which takes into account the
rearrangement of their electron densities as a result of the mutual
interaction, but in a perturbative way.

\subsubsection{The short-range binary potential}

The \textit{short-range } potential that represents the static interaction
of a neutral atom $A$, of nuclear charge $Z_{A}$ and electron density $%
n_{A}=n_{A}(\mathbf{r})$ \cite{Clementi-Roetti}, with an \textit{onion }$O$,%
\textit{\ }of nuclear charge $Z_{O}$ \ ($Z_{F_{@}^{-}}=9$ and $%
Z_{Li_{@}^{+}}=3$) and electron density $n_{O}=n_{O}(\mathbf{r})$, can be
approximated as a sum of four terms\ \cite{kimgordon74,ParrYang}:%
\begin{equation}
V_{AO}^{(short)}(R)=V_{e}(R)+V_{k}(R)+V_{x}(R)+V_{c}(R),  \label{110}
\end{equation}%
where $\mathbf{R}$ is the internuclear vector and $V_{e}$, $V_{k}$, $V_{x}$,
and $V_{c}$ are the electrostatic, kinetic, exchange and correlation
potentials, respectively. Due to the spherical symmetry of the interacting
partners, these partial potentials depend only on $R=\left\vert \mathbf{R}%
\right\vert $.

As explained in Sec. II A, the electron densities $n_{A}$ and $n_{O}$ are
here obtained from full HF calculations. Then, our task is to use them to
build the partial potentials involved in Eq. (\ref{110}). The first term of
Eq. (\ref{110}) represents the well-known electrostatic interaction, which
reads%
\begin{eqnarray}
V_{e}(R) &=&\frac{Z_{A}Z_{O}}{R}-\int d\mathbf{r}^{\prime }\ \frac{%
Z_{A}n_{O}(\mathbf{r})}{|\mathbf{r}-\mathbf{R}|}-\int d\mathbf{r}\ \frac{%
Z_{O}\ n_{A}(\mathbf{r}^{\prime })}{|\mathbf{r}^{\prime }+\mathbf{R}|}
\notag \\
&+&\iint d\mathbf{r}d\mathbf{r}^{\prime }\ \frac{n_{A}(\mathbf{r}^{\prime
})n_{O}(\mathbf{r})}{|\mathbf{R}+\mathbf{r}^{\prime }-\mathbf{r}|},
\label{120}
\end{eqnarray}%
while the remaining terms - $V_{k}$, $V_{x}$, and $V_{c}$ - can be derived
as \cite{ParrYang}:%
\begin{equation}
V_{j}(R)=E_{j}\left[ n_{tot}(\mathbf{R})\right] -E_{j}\left[ n_{O}\right]
-E_{j}\left[ n_{A}\right] ,\quad \text{for \ }j=k,x,c,  \label{140}
\end{equation}%
by assuming that the total electron density of the atom-\textit{onion}
system at a given internuclear separation $\mathbf{R}$ is given by \cite%
{ParrYang}
\begin{equation}
n_{tot}(\mathbf{R})=n_{O}(\mathbf{r})+n_{A}\left( \mathbf{r-R}\right) .
\label{130}
\end{equation}%
In Eq. (\ref{140}), the functionals $E_{j}\left[ n\right] $ for $j=k,x,c$
represent the kinetic, exchange and correlation energies, respectively,
depending not only on the local electron density $n(\mathbf{r})$, but also
on non-local magnitudes, like the gradient and the Laplacian\ of the
electron density.

In previous articles \cite{Gravielle2009,Gravielle2011} we used the
spin-restricted LDA to evaluate $E_{k}$ and $E_{x}\ $\ ($E_{c}$ was
neglected)$.$ In this article we do a quality leap by introducing non-local
approximations in terms of $\nabla n$ and\ $\nabla ^{2}n$, which allows us
to obtain\ more accurate values, but without losing the simplicity of Eq. (%
\ref{140}). For the exchange energy, $j=x$, we use the well established
spin-dependent Becke (B) approximation given by Eq. (8) of Ref. \cite%
{becke88}:%
\begin{equation}
E_{x}^{(B)}\left[ n\right] =c_{x}\int d\mathbf{r}\ n(\mathbf{r})^{4/3}\left(
1+\beta G(\mathbf{r})\right) ,  \label{x-becke}
\end{equation}%
where
\begin{equation}
G(\mathbf{r})=\frac{g(\mathbf{r})^{2}}{1+\gamma g(\mathbf{r})\sinh ^{-1}%
\left[ g(\mathbf{r})\right] },  \label{Gbecke}
\end{equation}%
with
\begin{equation}
g(\mathbf{r})=\left\vert \mathbf{\nabla }_{\mathbf{r}}n(\mathbf{r}%
)\right\vert /n(\mathbf{r})^{4/3},  \label{ggrad}
\end{equation}%
and $c_{x}$, $\beta $ and $\gamma $ are constants \cite{becke88}.
Accordingly, for the kinetic term, $j=k$, we use the Lee-Lee-Parr (LLP)
approach given by Eq. (7) of \ Ref. \ \cite{LLP91}, which can be considered
in a level equivalent to the B exchange expression since in terms of the
Density Functional Theory it is called "conjointness":

\begin{equation}
E_{k}^{(LLP)}\left[ n\right] =c_{k}\int d\mathbf{r}\ n(\mathbf{r}%
)^{5/3}\left( 1+\alpha G(\mathbf{r})\right) ,  \label{k-LLP}
\end{equation}%
where $c_{k}$ and $\alpha $ are constants \cite{LLP91}.

For the correlation energy, $j=c$, we use the celebrated Lee-Yang-Parr (LYP)
approximation, given by Eqs. (21) and (22) of Ref. \cite{LYP88}, which are
valid for closed- and open- shell atoms, respectively, \ also including $%
\nabla ^{2}n$. \ Hence, our full approximation for kinetic, exchange and
correlation terms should be called with the long acronym LLPB3LYP that means
Lee-Lee-Parr+Becke+3-coefficient-Lee-Yang-Parr \cite{Burke2012}. \

For the two different \textit{onions }- $F_{@}^{-}\ $\ or $Li_{@}^{+}$ - in
Table II we display : (\textit{i})\ the total energy calculated from HF \ ($%
E_{tot}^{(HF)}$); (\textit{ii}) the kinetic energy\ calculated from HF ($%
E_{k}^{(HF)}$),$\ $compared with values derived from the LDA \cite{ParrYang}
($E_{k}^{(LDA)}$) and from the LLP approximation as given by Eq. (\ref{k-LLP}%
) \cite{LLP91} ($E_{k}^{(LLP)}$ ) ; (\textit{iii}) the exchange energy
calculated from HF ($E_{x}^{\left( HF\right) }$), compared with values
derived from the LDA \cite{ParrYang} ($E_{x}^{\left( LDA\right) }$)\ and
from the B approach as given by Eq. (\ref{x-becke}) \cite{becke88} ($%
E_{x}^{(B)}$ ) ; and (\textit{iv}) the correlation energy evaluated by using
the LYP model \cite{LYP88}$\ $($E_{c}^{(LYP)}$). For both \textit{onions},
the kinetic and exchange energies derived with the functionals of Eqs. (\ref%
{k-LLP}) and (\ref{x-becke}), respectively, are in better agreement with the
corresponding HF values than the ones obtained from the LDA. Also the total
energies obtained including the correlation term $E_{c}^{(LYP)}$ are close
to the total HF values. In this regard, it is important to remind that the
Virial theorem does \textit{not} hold for this case because we are not
dealing with a central Coulomb potential.

\begin{figure}[tbp]
\includegraphics[width=0.5\textwidth]{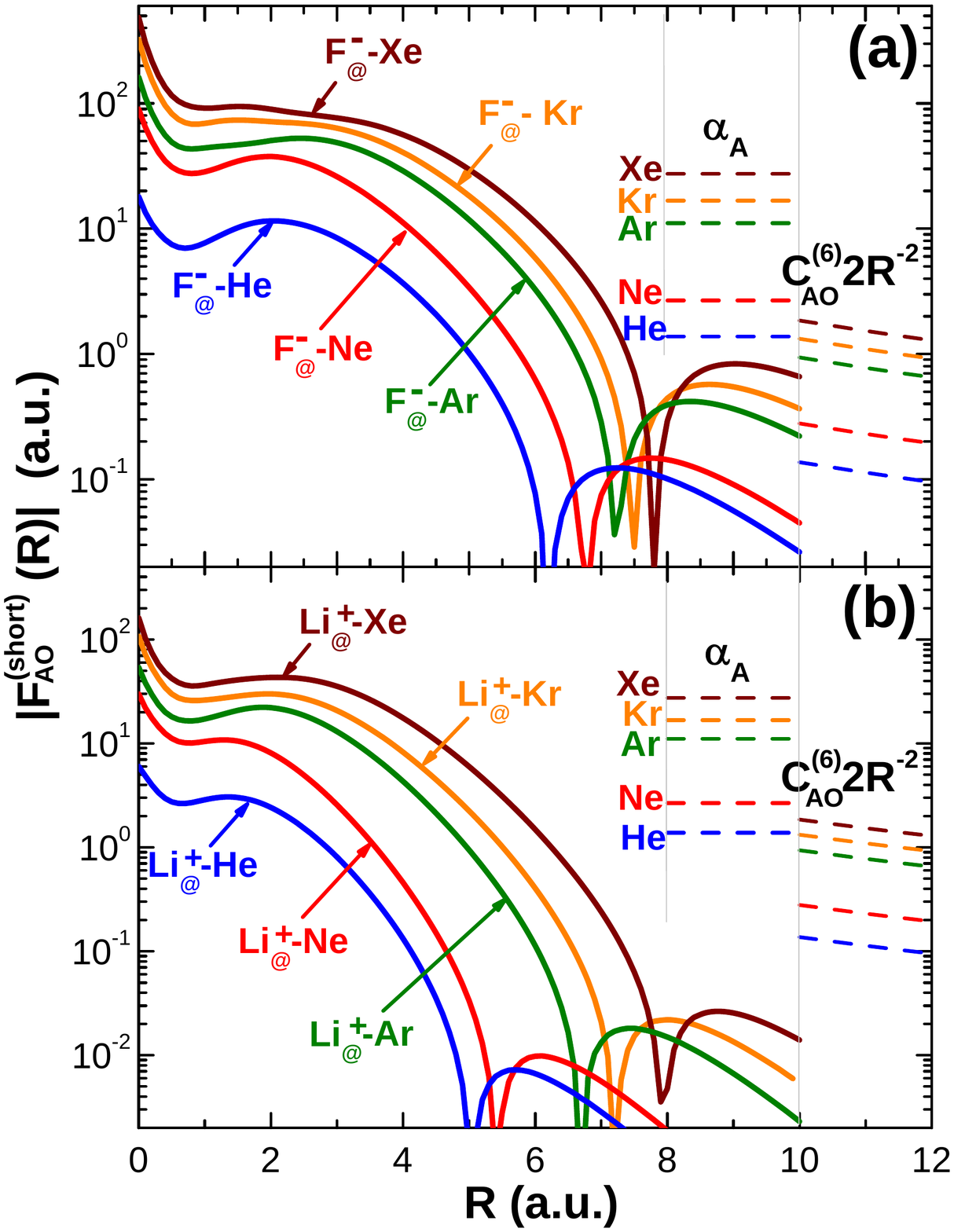}
\caption{(Color online) Binary interatomic potential, as a function of the
internuclear distance $R$, for different closed-shell atoms. (a) Absolute
value of $F_{AO}^{(short)}(R)$, given by Eq. (\protect\ref{150}), for the
interaction of different closed-shell atoms with $F_{@}^{-}$ anions. (b)
Analogous to (a) \ for the interaction with $Li_{@}^{+}$ cations. In both
panels, LLPB3LYP results for different atom-\textit{onion }pairs are plotted
with different colors. Projectile dipole polarizabilities, $\protect\alpha %
_{A}$, and van der Walls contributions, $C_{AO}^{(6)}2R^{-2}$, are displayed
in the ranges $8$ a.u.$\leq R\leq 10$ a.u. and $10$ a.u.$\leq R\leq 12$
a.u., respectively, as explained in the text.}
\label{interatomic1}
\end{figure}

\begin{figure}[tbp]
\includegraphics[width=0.5\textwidth]{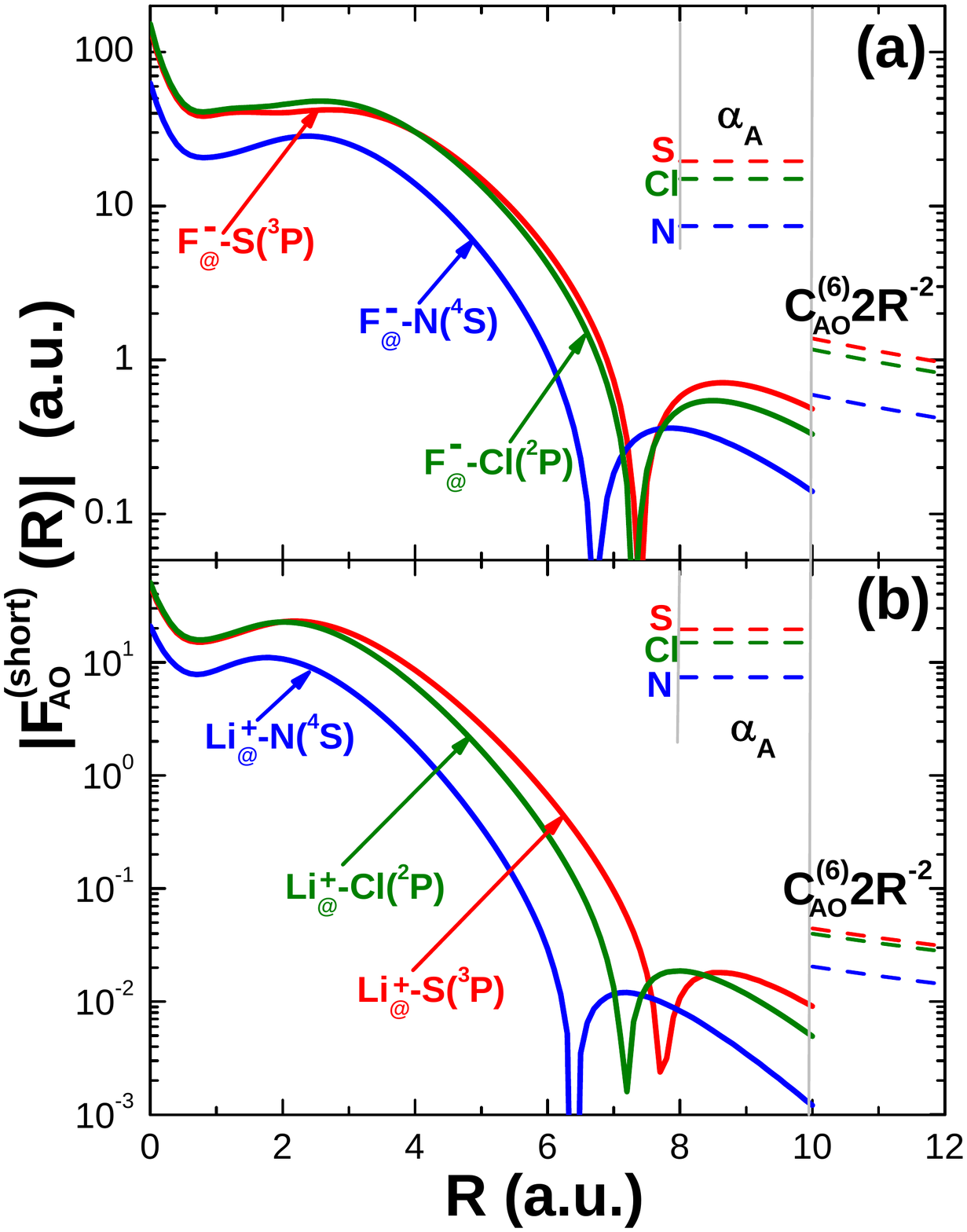}
\caption{(Color online) Analogous to Fig. 2 for the binary interaction of
open-shell atoms - N($^{4}S$), S($^{3}P$), and Cl($^{2}P$) - with: (a) $%
F_{@}^{-}$ anions and (b) $Li_{@}^{+}$ cations.}
\label{interatomic2}
\end{figure}

Results of our LLPB3LYP approximation for the short-range binary potentials
corresponding to $F_{@}^{-}$ and $Li_{@}^{+}\ $interacting with closed-shell
atoms - He, Ne, Ar, Kr, Xe- are displayed in Fig. \ref{interatomic1}. In
turn, in Fig. \ref{interatomic2} we focus on projectiles having open outer
shells - N($^{4}S$), S($^{3}P$) and Cl($^{2}P$) - which are ferromagnetic
atoms corresponding to the so-called \textit{unrestricted}$\mathit{\ }$spin
cases. For these latter projectiles, as well as for He, Ne and Kr, there are
experimental data of rainbow and/or FAD maxima available in the literature
\cite{Gravielle2011,Gravielle2013}. In Figs. 2 and 3, in order to analyze
straightforwardly the asymptotic limits of the short-range potentials,
results are displayed by means of the function%
\begin{equation}
F_{AO}^{(short)}(R)=V_{AO}^{(short)}(R)R(1+2R^{3}),  \label{150}
\end{equation}%
which makes evident the behavior at short and long distances. At the origin $%
F_{AO}^{(short)}(0)$ $=\ Z_{A}Z_{O}$, indicating that the internuclear atom-%
\textit{onion} interaction, given by the first term of Eq. (\ref{120}),
provides the main contribution to the binary potential $V_{AO}^{(short)}$
for small $R$ values. At large distances, instead, $F_{AO}^{(short)}(R)$
tends as $V_{AO}^{(short)}(R)2R^{4}$, which competes directly with the
polarizability of the impinging atom $\alpha _{A}\ $(see Eqs. (\ref{200})
and (\ref{201}) below). Besides, as in these figures we are plotting the
absolute value of $F_{AO}^{(short)}(R)$, its sign must be indicated: at
short distances $V_{AO}^{(short)}$ is always positive due to the static and
kinetic contributions, while at large distances $V_{AO}^{(short)}$ is
negative as a consequence of the preponderance of the exchange and
correlation energies.

\subsubsection{The long-range binary potential}

By \textit{long-range} binary potential we mean the potential produced by
the rearrangement of the electron densities of the interacting partners, \
also known as dispersive force, which dominates the long-distance behavior
of the atom-\textit{onion} interaction. Within a perturbative treatment, the
long- range binary potential for the system composed by a target \textit{%
onion} $O$, with $O=$ $F_{@}^{-}\ $or $O=$ $Li_{@}^{+}$, and an incident
neutral atom $A$ can be expanded as a power series of $\ R$, reading \cite%
{tang84,Meyer2015}:%
\begin{equation}
V_{AO}^{(long)}(R)\rightarrow -\frac{C_{AO}^{(4)}}{R^{4}}-\frac{C_{AO}^{(6)}%
}{R^{6}}-\frac{C_{AO}^{(8)}}{R^{8}}-...,  \label{200}
\end{equation}%
where $R$ is again the internuclear distance. The coefficient of first term
of Eq. (\ref{200}) is expressed as

\begin{equation}
\ C_{AO}^{(4)}=\frac{\alpha _{A}}{2},  \label{201}
\end{equation}%
where $\alpha _{A}$ is the static polarizability of the atom $\ A$. This
term is associated with the dipole momentum induced on the projectile by the
target \textit{onion} $O$, reflecting the contribution of the projectile
polarization. In Table III \ we list the values of the static
polarizabilities for the considered projectiles, as extracted from the
bibliography \cite{mitroy10,chu04}. Furthermore, to compare the contribution
of this term with the asymptotic limit of $V_{O}^{(short)}$, the \ $\alpha
_{A}$values are also displayed in Figs. 2 and 3 considering the range $%
R=8-10 $ a.u. where the dipolar term results relevant.

It is also interesting to investigate the following term of the expansion of
Eq. (\ref{200}), which is governed by the coefficient $C_{AO}^{(6)}$ related
to vdW forces. The value of $C_{AO}^{(6)}$ can be estimated by using the
formula of\ Slater-Kirkwood \cite{kirkwood31} as%
\begin{equation}
C_{AO}^{(6)}=\frac{3}{2}\frac{\alpha _{A}\alpha _{O}}{\left( \sqrt{\alpha
_{A}/N_{A}}+\sqrt{\alpha _{O}/N_{O}}\right) },  \label{195}
\end{equation}%
where $\alpha _{A}$ and $\alpha _{O}$ are the static polarizabilities of the
atom and the \textit{onion}, respectively, and $N_{A}$ and $N_{O}$ are the
numbers of the corresponding active electrons (i.e., the external ones).
Both magnitudes are well known for atomic projectiles: The atomic
polarizabilities $\alpha _{A}$ are given in Table III, while the $N_{A}$
values can be calculated from the homonuclear coefficients $C_{AA}^{(6)}\ $%
\cite{chu04} in the usual way (see Eq. (2) of Ref. \cite{koutselos86}). But
for \textit{onions} the values of $\alpha _{O}$ and $N_{O}$ must be
specifically determined as explained in the Appendix A.

Using the \textit{onion} values given by Eqs. (\ref{510}) and (\ref{520}),
together with the recommended atomic parameters \cite{mitroy10,chu04},
listed in Table III, we obtain the $C_{AO}^{(6)}$ values also tabulated in
the same table for the different atom-\textit{onion }systems. \ For He
atoms, our $C_{AO}^{(6)}$ coefficients are very close to the ones by Celli
\textit{et al.} \cite{celli85}, \ obtained by fitting experiments of helium
atoms bound to a LiF surface. Moreover, from the $C_{AO}^{(6)}$ \ values of
Table III we can evaluate the vdW contribution to the function
\begin{equation}
F_{AO}^{(long)}(R)=V_{AO}^{(long)}(R)2R^{4},  \label{R-long}
\end{equation}%
which reads $C_{AO}^{(6)}2R^{-2}$, also shown in Figs. 2 and 3 for $R\geq 10$
a.u. By comparing this contribution with $F_{AO}^{(short)}$ and with the
projectile polarizability, we are able to estimate that vdW forces affect
binary interatomic potentials only at very long distances, larger than $10$
a.u.

\begin{figure}[tbp]
\includegraphics[width=0.5\textwidth]{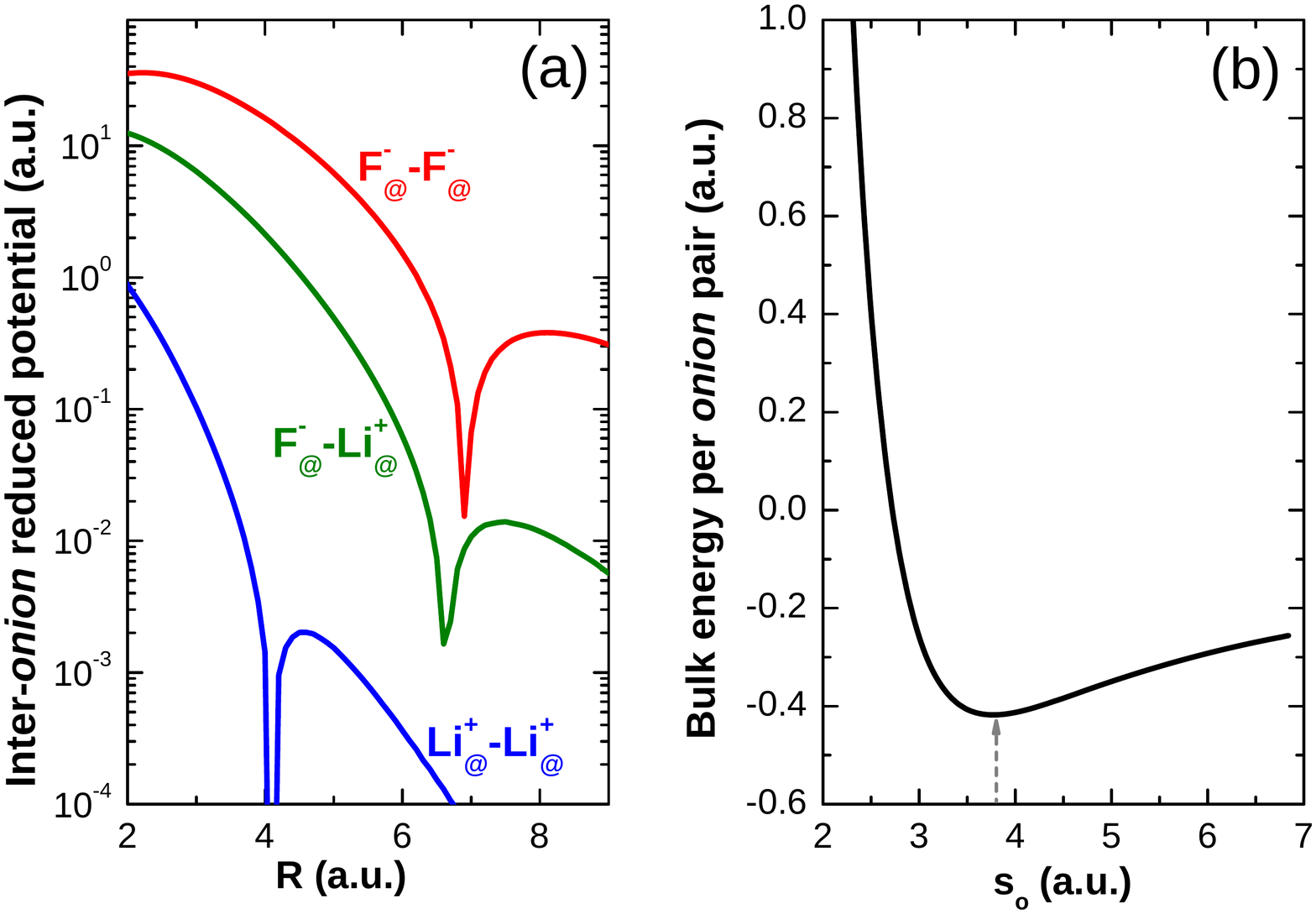}
\caption{(Color online) (a) Inter-\textit{onion} reduced potential (absolute
value), as a function of the internuclear distance $R$, for the following
\textit{onion}- pairs: $F_{@}^{-}-F_{@}^{-}$, $F_{@}^{-}-Li_{@}^{+}$, $%
Li_{@}^{+}-Li_{@}^{+}$. Such reduced potentials were derived within the
proposed potential model by multiplying by $R(1+2R^{3})$ (analogous to Eq. (%
\protect\ref{150})), after extracting the asymptotic Coulomb interaction.
(b) Energy per \textit{onion}- pair at the bulk, as a function of the
nearest-neighbor internuclear distance $s_{o}$. The vertical arrow indicates
the equilibrium position.}
\label{Xa}
\end{figure}

Finally, before tackling the evaluation of the total atom-surface potential,
it is interesting to use the same potential model to address the study of
the inter-\textit{onion} $F_{@}^{-}-F_{@}^{-}$, $F_{@}^{-}-Li_{@}^{+}$ and $%
Li_{@}^{+}-Li_{@}^{+}$ potentials, shown in Fig. \ref{Xa} (a). From these
potentials we evaluate the total energy per \textit{onion}-pair at the bulk
\cite{AshcroftMermin}, which is displayed in Fig. \ref{Xa} (b) as a function
of the nearest-neighbor \textit{onion} distance $s_{o}$. We can see that the
curve of Fig. \ref{Xa} (b) presents a minimum around $s_{o}=3.8$ a.u., which
is in very good agreement with the nearest-neighbor distance corresponding
to the real crystal, i.e. $s_{o}=a/2$.

\section{TOTAL\ ATOM-SURFACE\ POTENTIAL\ }

By using the short- and long- range binary potentials introduced in Sec. II,
we proceed to build the total atom-surface potential $W(\mathbf{R}_{A})$ for
an atom $A$ interacting with a LiF(001) surface. It reads:%
\begin{equation}
W(\mathbf{R}_{A})=W^{(short)}(\mathbf{R}_{A})+W^{(long)}(\mathbf{R}_{A}),
\label{300}
\end{equation}%
where $\mathbf{R}_{A}$ denotes the position of $A$ with respect to origin of
the frame of reference, placed on a given ionic\ center of the topmost
atomic layer (in our case, an \ $F_{@}^{-}$ -site) and\ $W^{(short)}$ ($%
W^{(long)}$) represents the short- (long-) range contribution to the total
atom-surface potential. \

The term $W^{(short)}(\mathbf{R}_{A})$ is expressed as the sum of the binary
short-range potentials given by Eq. (\ref{110}) as:%
\begin{equation}
W^{(short)}(\mathbf{R}_{A})=\sum\limits_{i}e_{i}V_{AO_{i}}^{(short)}(\mathbf{%
\rho }_{i}),  \label{310}
\end{equation}%
where $\mathbf{\rho }_{i}=\mathbf{R}_{A}-\mathbf{R}_{i}$, with $\mathbf{R}%
_{i}$ being the position vector of the \textit{onion} labeled with index $i$
($O_{i}$), and the factor $e_{i}$ describes the Evjen caging, that is, $%
e_{i}=1$ except for \textit{onions} placed at the limiting surface\ ($%
e_{i}=1/2$), at the arista$\ $($e_{i}=1/4$)$\ $ or at the vertex ($e_{i}=1/8$%
)\ of the crystal sample. This caging warranties the Coulomb neutrality of
the considered portion of crystal.\ In all our calculations, the sum on $i\ $%
includes $11\times 11\times 4=484\ $\ crystal sites. In addition, we
considered a surface rumpling with an outward (inward) shift of the
positions of the topmost $F_{@}^{-}$ \ ( $Li_{@}^{+}$) \textit{onions},
relative to the unreconstructed surface, of $0.046$ a.u., as extracted from
the \textit{ab-initio} calculation of Ref. \cite{Schuller2009}. In Fig. \ref%
{shortrange} we plot the short-range potential $W^{(short)}$ for the atoms
investigated in this work, as a function of the normal distance to the
surface, considering a position on top of an $F^{-}$ site.

In contrast with the short-range contribution, the long-range interaction $%
W^{(long)}(\mathbf{R}_{A})$ \textit{cannot} be obtained by simply adding the
corresponding binary potentials. The total long-range potential is here
split in two terms: $\ $%
\begin{equation}
W^{(long)}(\mathbf{R}_{A})=U^{(dip)}(\mathbf{R}_{A})+U^{(vdW)}(\mathbf{R}%
_{A}),\   \label{315}
\end{equation}%
where $U^{(dip)}$ and $U^{(vdW)}$\ \ correspond to the dipole and vdW
contributions, which are associated with the first and second term of \ Eq. (%
\ref{200}), respectively. We stress that each of these contributions is not
pairwise additive.

The dipole potential $U^{(dip)}(\mathbf{R}_{A})$ depends on the total
electric field produced by the different ionic centers of the crystal,
evaluated at the position of the atom $A$. It reads
\begin{equation}
U^{(dip)}(\mathbf{R}_{A})=-\frac{\alpha _{A}}{2}\left\vert \sum\limits_{i}%
\mathbf{E}_{i}(\mathbf{R}_{A})\right\vert ^{2},  \label{320}
\end{equation}%
where
\begin{equation}
\mathbf{E}_{i}(\mathbf{R}_{A})=f_{i}(\rho _{i})Z_{O_{i}}^{(\infty )}\frac{%
\widehat{\mathbf{\rho }}_{i}}{\rho _{i}^{2}}  \label{330}
\end{equation}%
is the electric field produced by the asymptotic charge of the \textit{onion}
$O_{i}$, with $Z_{O_{i}}^{(\infty )}=-1$ ($+1$) for $O_{i}=F_{@}^{-}$ ($%
Li_{@}^{+}$),\textbf{\ }$\widehat{\mathbf{\rho }}_{i}=\mathbf{\rho }%
_{i}/\rho _{i}$, and $f_{i}(\rho _{i})$ is a screening factor that avoids
the divergence of this electric field at the origin. In this work the
function $f_{i}$ was evaluated taking into account information about the
physics of adatoms, as explained in the Appendix B. Noteworthily, the
projectile polarization term given by Eq. (\ref{320}) strongly affects FAD
spectra for incidence along the $\langle 110\rangle $ channel \cite%
{Gravielle2008,Gravielle2011}. But for incidence along the $\langle
100\rangle $ direction, the alternation of the opposite effective Coulomb
charges of the $F_{@}^{-}\ $and $Li_{@}^{+}$ \textit{onions} along the
channel makes the polarization effect negligible \cite%
{Gravielle2008,Gravielle2011}.

In the case of the vdW contribution $U^{(vdW)}(\mathbf{R}_{A})$, for the
sake of simplicity we evaluate it at a position on top of an F$^{-}$ site;
that is, at $\mathbf{R}_{A}=z_{A}\ \widehat{\mathbf{z}}$, where the versor $%
\widehat{\mathbf{z}}$ is oriented normal to the surface, aiming towards the
vacuum region. In the most simple model, far from the surface $%
U^{(vdW)}(z_{A}\ \widehat{\mathbf{z}})$ can be derived as the superposition
of the binary vdW contributions (second term of Eq. (\ref{200})) \ produced
by a continuous distribution of \textit{onions}, reading
\begin{eqnarray}
U^{(vdW)}(z_{A}\ \widehat{\mathbf{z}}) &\rightarrow &\sum -\delta
_{v}\int\limits_{-\infty }^{0}dz_{o}\iint\limits_{-\infty }^{+\infty
}dx_{o}dy_{o}  \notag \\
&\times &\frac{\ \left( C_{AF_{@}^{-}}^{(6)}+C_{ALi_{@}^{+}}^{(6)}\right) \
\ }{\left[ (z_{A}-z_{o})^{2}+x_{o}^{2}+y_{o}^{2}\right] ^{3}}  \notag \\
&\simeq &-\frac{D^{^{(vdW)}}}{(z_{A}-\overline{d})^{3}}\text{ \ \ \ \ \ \ \
\ \ \ \ \ for }z_{A}\rightarrow +\infty ,  \notag \\
&&  \label{430}
\end{eqnarray}%
where
\begin{equation}
D^{^{(vdW)}}=\frac{\pi }{6}\delta _{v}\left(
C_{AF_{@}^{-}}^{(6)}+C_{ALi_{@}^{+}}^{(6)}\right) ,  \label{431}
\end{equation}%
$\delta _{v}=4/a^{3}$ is the volume density of each \textit{onion } and $%
\overline{d}$ is a reference distance that does not have direct relation
with the equilibrium position.

\ In relation to Eq. (\ref{430}), we must mention that it does not take into
account the fact that the different dipoles induced in the crystal by the
projectile interaction screen each other \cite{celli85}. For this reason, $%
D^{^{(vdW)}}$ values derived from Eq. (\ref{431}) should be considered as an
upper limit because they are expected to be higher than those obtained from
the approach by Lifshitz\ \textit{et al}. \cite{lifshitz56,vidali91}, which
includes the proper screening. Remarkably, we observe that the $z_{A}^{-3}\ $%
dependence given by Eq. (\ref{430}) , which gave rise to the famous
potential $V_{9-3}$ \cite{vidali91}, starts to dominate at distances farther
than $10$ a.u. This fact makes the influence of $U^{(vdW)}$ negligible for
normal energies higher than $0.2$ eV, like the ones considered in this
article, where closest distances smaller than $5.3$ a.u. are reached by the
impinging atoms.

Concerning the importance of the vdW contribution, we should draw the
attention to H projectiles, for which a completely different situation is
observed. In the case of FAD for H on LiF(001), a non-negligible role of vdW
interactions was recently reported in the low-to-intermediate normal energy
regime \cite{Bocan2016}. Such a noticeable vdW effect is compatible with the
relatively high $C_{AO}^{(6)}$ values for hydrogen atoms, in comparison with
the corresponding short-range potentials, obtained within our \textit{onion}
model: $C_{HF_{@}^{-}}^{(6)}=15.1$ a.u. and $C_{HLi_{@}^{+}}^{(6)}=0.47$ a.u.

\section{COMPARISON WITH GRAZING\ INCIDENCE\ EXPERIMENTS}

With the goal of checking the quality of the proposed surface
potential we use the potential model \ within the framework of the
SIVR approximation in order to evaluate final projectile
distributions for swift atoms grazingly impinging on LiF(001) along
low-indexed crystallographic channels. The SIVR approach
\cite{Gravielle2014,Gravielle2015} is a semiquantum method that
provides a clear representation of the main physical mechanisms
involved in FAD processes, describing them in terms of classical
trajectories through
the Feynman path integral formulation of quantum mechanics \cite{Miller2001}%
. It incorporates an adequate description of classically forbidden
transitions on the dark side of the rainbow angle, providing reliable FAD
patterns along the whole angular range \cite{Gravielle2014,Gravielle2015}.

Under axial incidence conditions \cite{Winter2011}, like the ones considered
here, the angular distribution of scattered projectiles lays on an annulus
given by $\varphi _{f}^{2}+\theta _{f}^{2}\approx \theta _{i}^{2}$, where $%
\theta _{f}$ ($\theta _{i}$) is the final (initial) polar angle, measured
with respect to the surface, and $\varphi _{f}$ \ is the azimuthal exit
angle\ measured with respect to the incidence direction in the surface
plane, as shown in the inset of Fig. \ref{rainbow}. The typical FAD
distribution displayed in such an inset presents maxima symmetrically placed
with respect to $\varphi _{f}$ $=0$, which are associated with rainbow and
supernumerary rainbow peaks. The outermost maxima of the spectrum are
produced by rainbow scattering, having a classical explanation, while the
inner peaks are related to supernumerary rainbows, being originated by
quantum interference among trajectories with starting positions placed
inside \textit{one} reduced unit cell of the crystal surface. Furthermore,
for LiF surfaces when the final projectile distribution is plotted as a
function of the deflection angle $\Theta $, defined as $\Theta =\arctan
(\varphi _{f}\ /\theta _{f})$, both the position and intensity of the peaks
become completely governed by the normal energy $E_{\perp }=E\sin ^{2}\theta
_{i}$, which is related to the motion in the plane perpendicular to the
incident channel, with $E$ being the impact energy. \ The angular positions
of the peaks are strongly affected by the corrugation of the atom-surface
potential across the axial direction, making possible to probe the potential
model for different distances to the surface by varying the $E_{\perp }$
value. Notice that for low normal energies, both rainbow and supernumerary
rainbow peaks are present in FAD spectra, but when $E_{\perp }$ increases
supernumerary rainbows start to blur out, and for large energies only
rainbow structures are observed in the projectile distributions.

To show an overall picture of performance of the proposed potential model,
in Fig. \ref{rainbow} we plot the deflection angle corresponding to the
rainbow maximum, $\Theta _{\mathrm{rb}}$, as a function of $E_{\perp }$, for
closed- shell (Ne and Kr) and open- shell (S and Cl) atoms grazingly
colliding with LiF(001) along two different channels: $\langle 110\rangle $
and \ $\langle 100\rangle $. In the figure, $\Theta _{\mathrm{rb}}$ values
obtained from the proposed pairwise potential model are compared with
available experimental data \cite{Gravielle2011,Gravielle2013} considering a
wide normal energy range that covers from $0.2$ to $80$ eV. Note that for
high normal energies, the angular positions of rainbow maxima derived from
the SIVR method agree with those obtained from classical simulations due to
the classical origin of rainbow scattering. In the SIVR calculations, the
atom-surface potential was evaluated as the sum of a short-range
contribution, derived from the LLPB3LYP approach, and a \ long-range
contribution accounting for projectile polarization through the dipolar term
given by Eq. (\ref{320}). The vdW contribution was not included in the
simulations because it was estimated as negligible, as explained in Sec. III.

From Fig. \ref{rainbow} we observe that for closed-shell projectiles, our
potential model yields rainbow angles in very good accord with the
experimental data in the whole $E_{\perp }$-range. But for S and Cl
projectiles, both open-shell atoms, the theoretical $\Theta _{\mathrm{rb}}$
values agree with the experimental ones for \ $E_{\perp }$ up to $60$ and $%
30 $ eV, respectively, running slightly below the experiments at higher
normal energies. This underestimation of rainbow deflection angles in the
high normal-energy region is associated with a less corrugated potential
energy surface, which affects projectiles that reach closer distances to the
surface, with maximum approach distances lower than $2.0$ a.u. In Fig. \ref%
{rainbow} we also investigate the influence of the correlation term, usually
left out in pairwise potential models, by plotting \ $\Theta _{\mathrm{rb}}$
values obtained by neglecting the correlation contribution $E_{c}$, that is,
by using the LLPB approximation, instead of the LLPB3LYP one, to represent
the short-range binary interactions. In all the spectra we found that
rainbow deflection angles derived from the LLPB approach are very close to
those obtained with LLPB3LYP, indicating a weak effect of the correlation
term on rainbow scattering.

Additionally, in the low energy region it is possible to make a more
exhaustive analysis of the surface potential by using FAD patterns. In Fig. %
\ref{FAD} angular positions of rainbow and supernumerary rainbow maxima,
obtained from SIVR simulations\ for Ne (closed-shell) and N (open-shell)
atoms impinging on LiF(001) along the $\langle 110\rangle $ direction, are
compared with experimental FAD data \cite{Gravielle2011,Winter-priv}, as a
function of $E_{\perp }$. \ These FAD maxima represent a sensitive test of
the corrugation of the potential across the channel. \ In this regard, we
have chosen the $\langle 110\rangle $ channel, instead\ of the $\langle
100\rangle $ one, because the former presents a higher corrugation of the
potential across it, producing consequently richer diffraction patterns,
with a larger number of supernumeraries. Notice that the energy range where
supernumerary maxima can be experimentally resolved depends on the
projectile mass \cite{Gravielle2011}, being smaller for Ne than for N
impact. For Ne projectiles, the SIVR approximation, using the proposed
pairwise additive potential, reproduces well the experimental positions of
rainbow and supernumerary rainbow maxima over the whole energy range. But
the experimental data are slightly underestimated when the correlation term
is neglected in the binary\ potentials. On the contrary, for N projectiles
the inclusion of the correlation term gives rise to an increase of the
deflection angles corresponding to rainbow and supernumerary rainbow peaks,
overestimating slightly the experimental values, which are better reproduced
when the LLPB model, without correlation, is used in the calculations.
Similar behavior was also observed in FAD spectra for He projectiles, case
investigated with the present pairwise potential model in Refs. \cite%
{Gravielle2014,Gravielle2015}.

\section{CONCLUSIONS}

We have derived a pairwise additive potential for neutral atoms interacting
with a LiF(001) surface by using state-of-the-art methods to describe binary
potentials in terms of the electron density. The model incorporates not only
non-local contributions of the electron density, but also the effect of the
ionic crystal lattice on the electron density around individual ionic sites
of the material, here named \textit{onions}. For closed-shell (He, Ne, Ar,
Kr, and Xe) and open-shell (N, S, and Cl) atoms, the short-range and long-
range contributions to the corresponding atom-\textit{onion} potentials were
analyzed as a function of the internuclear distance, comparing the relative
importance of each term.

The degree of accuracy \ of the proposed atom-surface potential model was
illustrated by contrasting theoretical angular positions of rainbow maxima
with experimental data for axial grazing scattering \ with normal energies
from $0.2$ to $80$ eV. In this energy range \ the vdW contribution to the
surface potential was estimated as negligible and consequently, it was not
included in the calculations. For two different incidence directions - $%
\langle 110\rangle $ and $\langle 100\rangle $ - and closed-shell
projectiles, the rainbow angles derived from the proposed potential were
found in excellent agreement with the experiments in the whole $E_{\perp }$-
range. For open-shell atoms, instead, an analogous accord was observed up to
intermediate $E_{\perp }$\ values, but \ in the high normal energy region
our theoretical results slightly underestimate the experiment. In addition,
the low $E_{\perp }$- region of the surface potential was deeply probed by
comparing theoretical and experimental angular positions of rainbow and
supernumerary rainbow maxima, which are produced by very sensitive FAD
processes. Experimental data for closed-shell Ne atoms as well as for
open-shell N atoms were fairly well reproduced by our surface potential.
Furthermore, in all the cases the correlation term was found to play a minor
role, its effect being appreciable for low $E_{\perp }$- values only.
Therefore, we conclude that our pairwise additive potential model can be
used with confidence for closed-shell atoms in the $0.2$ -$80$ eV normal
energy range and for open-shell atoms up to intermediate energies. It is
important to point out that such a potential model presents the advantage of
allowing one to extract straightforwardly information about the different
interaction mechanisms.

\begin{acknowledgments}
Experimental contribution from the group of Helmut Winter (Institut f\"{u}r
Physik, Humboldt-Universit\"{a}t zu Berlin) is gratefully acknowledged. The
authors thank financial support from CONICET, UBA, and ANPCyT of Argentina.
\end{acknowledgments}

%%%%%%%%%%%%%%%%%%%%%%%%%%%%%%%%%%%%%%%%%%%%%%%%%%%%%%%%%%%%%%%%%%%%%%%%%%%%%
\appendix

\section{\textit{Onion} polarizabilities}

The polarizabilities of the $F_{@}^{-}\ $and $Li_{@}^{+}$ \textit{onions},
namely, $\alpha _{F_{@}^{-}}$ and $\alpha _{Li_{@}^{+}}$, should slightly
differ from the polarizabilities of the \textit{free} ions , given by $%
\alpha _{\mathrm{F}^{-}}=10.6$ a.u. and $\alpha _{\mathrm{Li}^{+}}=0.188$
a.u. for F$^{-}$ and Li$^{+}$, respectively \cite{mitroy10,shannon06}. Then,
we use the fact that the polarizability is proportional to the volume \cite%
{Kannemann12} to estimate the \textit{onion }polarizabilities as%
\begin{eqnarray}
\alpha _{F_{@}^{-}} &\simeq &\frac{\left\langle r^{3}\right\rangle
_{F_{@}^{-}}}{\left\langle r^{3}\right\rangle _{\mathrm{F}^{-}}}\alpha _{%
\mathrm{F}^{-}}=8.69\text{ a.u.},  \notag \\
\alpha _{Li_{@}^{+}} &\simeq &\frac{\left\langle r^{3}\right\rangle
_{Li_{@}^{+}}}{\left\langle r^{3}\right\rangle _{_{\mathrm{Li}^{+}}}}\alpha
_{\mathrm{Li}^{+}}=0.191\text{ a.u.},  \label{510}
\end{eqnarray}%
where $\left\langle r^{3}\right\rangle _{j}$ is the mean volume of the ion $%
j $, with $j=$ F$^{-}$, Li$^{+}$, $F_{@}^{-}$, $Li_{@}^{+}$ \cite{volu-r}.

Concerning the number of active electrons of the \textit{onion}, we exploit
the rule of Ref. \cite{koutselos86} that shows that the same number of
active electrons can be applied to an entire isoelectronic sequence, no
matter the charge state, without affecting the accuracy of Eq. (\ref{195}).
Therefore, from Table III:
\begin{eqnarray}
N_{F_{@}^{-}} &=&N_{Ne}=3.59,  \notag \\
N_{Li_{@}^{+}} &=&N_{He}=1.36.  \label{520}
\end{eqnarray}

\section{Screening function derived from adatom eigenenergies}

In Eq. (\ref{330}) we have introduced a screening function $f_{i}(\rho _{i})$%
, extracted\ from Ref. \cite{tang84}, which avoids the divergence of the
electric field at the origin. It reads%
\begin{equation}
f_{i}(\rho _{i})=1-\exp (-\frac{\rho _{i}}{\mu _{i}})\left[
\sum\limits_{j=0}^{2}\frac{1}{j!}\left( \frac{\rho _{i}}{\mu _{i}}\right)
^{j}\right] ,  \label{f-screen}
\end{equation}%
where $\mu _{i}$ is a screening parameter defined as $\mu _{i}=\eta \left(
\sqrt{\left\langle r^{2}\right\rangle _{O_{i}}}+\sqrt{\left\langle
r^{2}\right\rangle _{A}}\right) $, with $\left\langle r^{2}\right\rangle
_{O_{i}}$and $\left\langle r^{2}\right\rangle _{A}$ being the mean square
radii corresponding to the outer shell of the \textit{onion} and the atom,
respectively, and $\eta $ is an external factor that is here calculated from
the physics of adatoms as follows.

Atom-surface potentials calculated from Eq. (\ref{300}) should support
adatoms in front of the LiF\ surface, that is, atoms weakly bound to the
surface\ \cite{vidali91}. In the case of physical absorption from a surface,
the binding potential is characterized by the depth of the attractive well $%
V_{b}$, the ground-state binding energy $E_{b}$, and the equilibrium
distance $d_{b}$ from the surface, so that around this position the
potential behaves as a harmonic oscillator. To compare with adatom values,
we introduce an averaged potential, without corrugation (planar), valid at a
large distance $d$ \ from the surface, as

\begin{equation}
W(\mathbf{R}_{A})\simeq \frac{1}{2}W(\mathbf{R}_{\mathrm{F}})+\frac{1}{2}W(%
\mathbf{R}_{\mathrm{Li}})+W(\mathbf{R}_{\mathrm{hole}})  \label{360}
\end{equation}%
where $\mathbf{R}_{A}=\mathbf{R}_{\mathrm{F}}=(0,0,d)$ is the position upon
the$\ F_{@}^{-}$ \textit{onion}, $\mathbf{R}_{\mathrm{Li}}=(0,a/2,d)$ is the
one upon $Li_{@}^{+},$ and $\ \mathbf{R}_{\mathrm{hole}}=(a/4,a/4,d)$ is the
position upon the center of the reduced unit cell (hole).

For He projectiles, using $\eta =0.7$ and neglecting any vdW contribution,
we found that $V_{b}=8.0\ $meV and $d_{b}=2.99\ $\AA , in close agreement
with the reported experimental data $8.5$ meV and $2.98$ \AA \thinspace ,
respectively \cite{vidali91}. The so-obtained potential $W(\mathbf{R}_{A})$
can be fitted with a Morse potential to give a binding energy $E_{b}=5$ meV
that compares quite well with the best estimate value $5.9$ meV\ \cite%
{vidali91}. Similarly, for Ne, using $\eta =0.72$, we found $V_{b}=12.7\ $%
meV and $E_{b}=10.9$ meV, while the reported data are $13.5$ and
$11.7$ meV, respectively \cite{vidali91}. For the remaining rare
gases: Ar, Kr, and Xe we use $\eta =0.67$,$\ 0.64$, and $0.62,$
producing $V_{b}=\ 69$,\ \ $93$, and $157$ meV, comparable with the
reported values $70\pm 10$,$\ 94.2$, and$\ 153$ meV, respectively
\cite{vidali91}. Therefore, the $\eta $ \ value seems to be rather
universal and situates around $0.7$, in accordance with our previous
findings \cite{Gravielle2011,Gravielle2013,Gravielle2014}. In this
work we used such a value for our SIVR simulations.

\bigskip

%%%%%%%%%%%%%%%%%%%%%%%%%%%%%%%%%%%%%%%%%%%%%%%%%%%%%%%%%%%%%%%%%%%%%%%%%%%%%
%\bibliographystyle{plain}
\bibliographystyle{unsrt}
\bibliography{potSdft-n}

\newpage

\begin{figure}[tbp]
\includegraphics[width=0.5\textwidth]{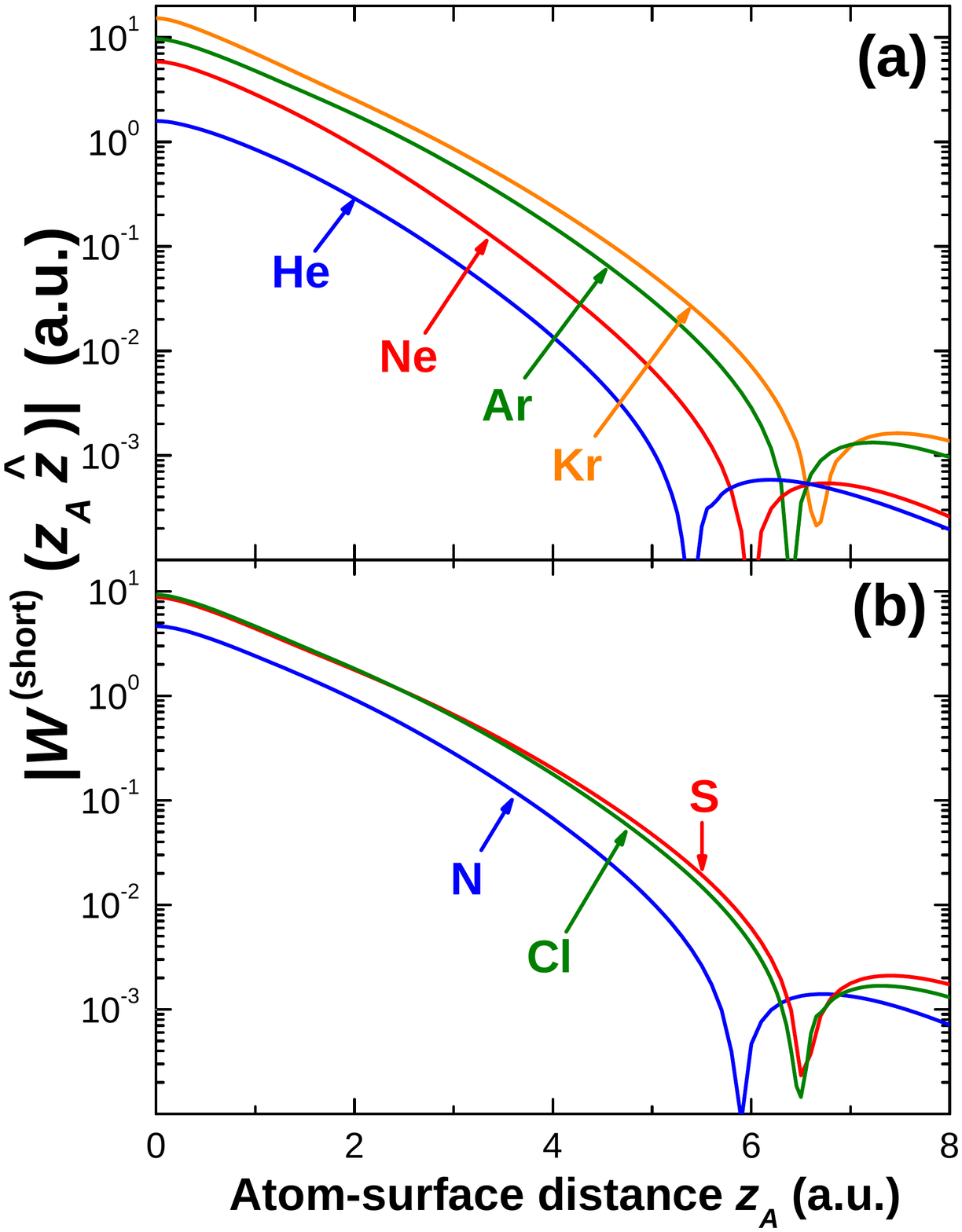}
\caption{(Color online) Absolute value of the short-range atom-surface
potential $W^{(short)}(z_{A}\widehat{\mathbf{z}})$, as defined by Eq. (%
\protect\ref{310}), as a function of the atom-surface distance $z_{A}$
measured on top of an F$^{-}$ site. Interactions with different (a)
closed-shell and (b) open-shell atoms are displayed with different colors.}
\label{shortrange}
\end{figure}

\begin{figure}[tbp]
\includegraphics[width=0.75\textwidth]{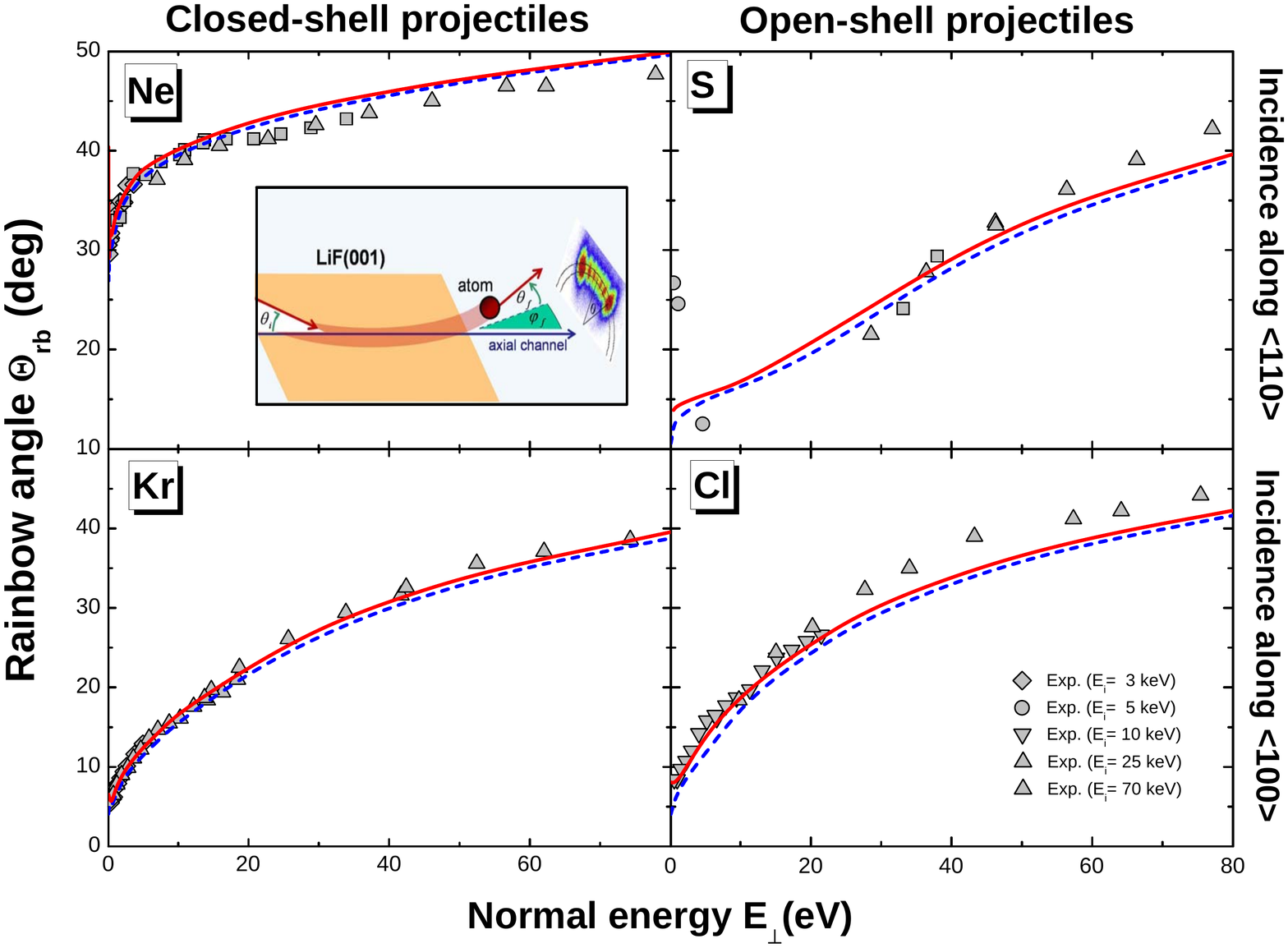}
\caption{(Color online) Rainbow deflection angle $\Theta _{\text{rb}}$, as a
function of the normal energy $E_{\perp }$, for closed-shell atoms - Ne and
Kr - in the left column, and for open-shell atoms - S and Cl - in the right
column. Panels in the upper and lower rows correspond to the incidence
directions $\langle 110\rangle $ and $\langle 100\rangle $, respectively.
Red solid line, results obtained from the proposed LLPB3LYP model; blue
dashed line, values derived from the LLPB model (neglecting the correlation
term); solid symbols, experimental data for different impact energies
extracted from Ref. \protect\cite{Gravielle2011,Gravielle2013}. Inset:
Depiction of the FAD processes.}
\label{rainbow}
\end{figure}

\begin{figure}[tbp]
\includegraphics[width=0.5\textwidth]{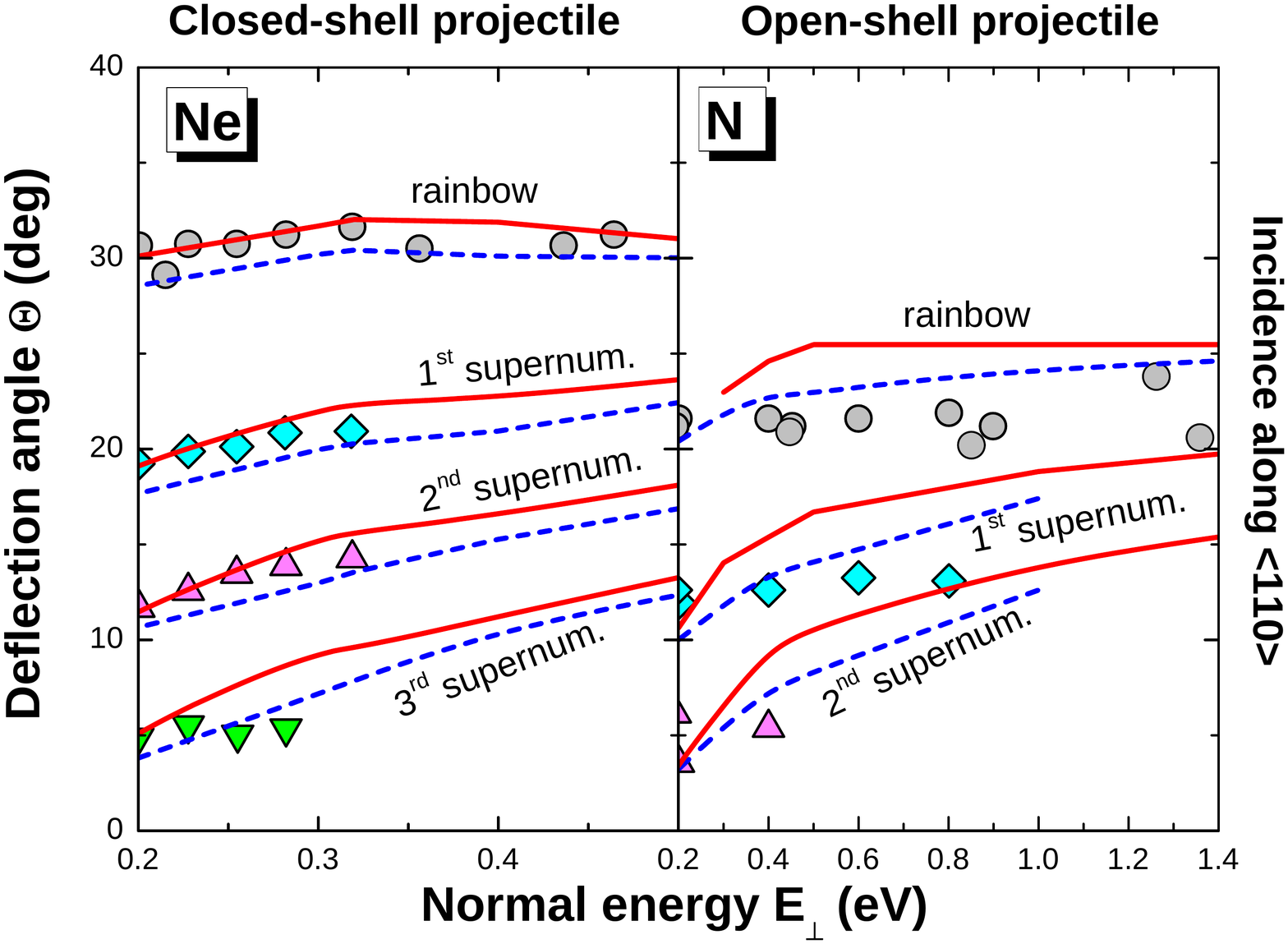}
\caption{(Color online) Deflection angles $\Theta $ corresponding to maxima
of FAD distributions, as a function of the normal energy $E_{\perp }$, for
(a) Ne (closed-shell) and (b) N (open-shell) atoms scattered along the $%
\langle 110\rangle $ direction. In both panels, red solid (blue dashed)
line, SIVR rainbow and supernumerary rainbow angles derived from the
LLPB3LYP (LLPB) models, including (without including) the correlation term.
Symbols: experimental data for rainbow (circles) and first (diamonds),
second (up triangles), and third (down triangles) supernumerary rainbow
angles, extracted from Refs. \protect\cite{Gravielle2011,Winter-priv}.}
\label{FAD}
\end{figure}

\begin{table*}[tph]
\centering%
\begin{tabularx}{0.65\textwidth}{|l|rrrrrr|rrr|}
\hline
$$
& $F_@^-(1s)$ & $F_@^-(2s)$ & $F_@^-(2p)$ & $F_@^-(3s)$ & $F_@^-(3p)$ &
$F_@^-(3d)$ & $Li_@^+(1s)$ & $Li_@^+(2s)$ & $Li_@^+(2p)$ \\
\tableline $E_{nl} $ & -26.17 & -1.447 & -0.5532 & -0.1593 & -0.0994
&
-0.0559 & -2.348 & -0.2249 & -0.1499 \\
$\langle r \rangle _{nl} $ & 0.1758 & 1.033 & 1.208 & 4.804 & 6.703
& 10.36 & 0.560 & 3.413 & 3.948 \\ \hline
\end{tabularx}
\caption{ HF bound energies and mean-radii for the considered \textit{onions}%
. All the values in atomic units.}
\label{Tab1}
\end{table*}

\begin{table*}[tph]
\centering
\begin{tabularx}{0.57\textwidth}{|l|r|rrr|rrr|r|}
\hline
$$
& $E_{\text{tot}}^{(HF)}$ & $E_k^{(HF)}$ & $E_k^{(LDA)}$ & $E_k^{(LLP)}$ & $E_x^{(HF)}$ &
$E_x^{(LDA)}$ & $E_x^{(B)}$ & $E_c^{(LYP)}$ \\
\tableline $F_@^-$ & -103.766 & 99.866 & 91.151 & 99.624 & -10.545 &
-9.449
& -10.423 & -0.363 \\
$Li_@^+$ & -6.3414 & 7.200 & 6.507 & 7.253 & -1.656 & -1.507 &
-1.640 & -0.0048 \\ \hline
\end{tabularx}
\caption{ Total $E_{tot}$, kinetic $E_{k}$, exchange $E_{x}$ and correlation
$E_{c}$ energies calculated with HF, LDA, LLP \protect\cite{LLP91} (Eq. (%
\protect\ref{k-LLP})) , B \protect\cite{becke88} (Eq. (\protect\ref{x-becke}%
)), and LYP \protect\cite{LYP88}, respectively, for the two \textit{onions}.
All the values in atomic units.}
\label{Tab2}
\end{table*}

\begin{table}[tph]
\centering
\begin{tabular}{|c|ccccccc|}
\hline
$A$ \ \  & \ \ $\alpha_A$ \ \  & \ \ $N_A$ \ \  & \ \ $C^{(6)}_{A-A}$ \ \  &
\ \ $C^{(6)}_{A-F_@^-}$ \ \  & \ \ $C^{(6)}_{A-Li_@^+}$\ \  &  &  \\ \hline
\ $\mathrm{{He} \ }$ & 1.38 & 1.36 & 1.42 & 7.02 & 0.29 &  &  \\
\ $\mathrm{{N} \ }$ & 7.40 & 2.57 & 24.2 & 29.7 & 1.02 &  &  \\
\ $\mathrm{{Ne} \ }$ & 2.67 & 3.59 & 6.20 & 14.4 & 0.62 &  &  \\
\ $\mathrm{{S} \ }$ & 19.6 & 4.24 & 134. & 68.9 & 2.22 &  &  \\
\ $\mathrm{{Cl} \ }$ & 15.0 & 4.71 & 94.6 & 58.5 & 1.99 &  &  \\
\ $\mathrm{{Ar} \ }$ & 11.1 & 5.36 & 64.2 & 48.3 & 1.75 &  &  \\
\ $\mathrm{{Kr} \ }$ & 16.7 & 6.45 & 130. & 68.8 & 2.41 &  &  \\
\ $\mathrm{{Xe} \ }$ & 27.3 & 5.96 & 261. & 96.2 & 3.11 &  &  \\ \hline
\end{tabular}%
\caption{ Dipole polarizability $\protect\alpha _{A}$, number of active
electrons $N_{A}$, and $C^{(6)}$ coefficient for the considered atoms ($A$)
\protect\cite{mitroy10,chu04,koutselos86} and atom-\textit{onion} pairs. All
the values in atomic units.}
\label{Tab3}
\end{table}

\end{document}